\documentclass[final]{svjour3}
\usepackage{graphicx}
\usepackage{rotating}
\usepackage{amssymb}
\usepackage{amsmath}
\usepackage{mathptmx}
\usepackage[square, numbers]{natbib}
\usepackage{xcolor}
\usepackage{comment}
\usepackage{bm}
\usepackage{siunitx}

\usepackage{multirow}
\makeatletter
\usepackage[symbol]{footmisc}
\usepackage[affil-it]{authblk}

\journalname{Journal of Low Temperature Physics}

\begin{document}

\newcommand{\hdblarrow}{H\makebox[0.9ex][l]{$\downdownarrows$}-}
\title{Small Aperture Telescopes for the Simons Observatory}

\author{Aamir M. Ali\textsuperscript{*,1}\kern-1.5pt
\and Shunsuke Adachi \textsuperscript{2}\kern-1.5pt
\and Kam Arnold \textsuperscript{3}\kern-1.5pt
\and Peter Ashton\textsuperscript{1,4,5}\kern-1.5pt
\and Andrew Bazarko \textsuperscript{6}\kern-1.5pt
\and Yuji Chinone \textsuperscript{1,5,7}\kern-1.5pt
\and Gabriele Coppi \textsuperscript{8}\kern-1.5pt
\and Lance Corbett \textsuperscript{1}\kern-1.5pt
\and Kevin D Crowley \textsuperscript{6}\kern-1.5pt
\and Kevin T Crowley \textsuperscript{1}\kern-1.5pt
\and Mark Devlin \textsuperscript{8}\kern-1.5pt
\and Simon Dicker \textsuperscript{8}\kern-1.5pt
\and Shannon Duff \textsuperscript{9}\kern-1.5pt
\and Chris Ellis \textsuperscript{3}\kern-1.5pt
\and Nicholas Galitzki \textsuperscript{3}\kern-1.5pt
\and Neil Goeckner-Wald \textsuperscript{10,11}\kern-1.5pt
\and Kathleen Harrington \textsuperscript{12}\kern-1.5pt
\and Erin Healy \textsuperscript{6}\kern-1.5pt
\and Charles A Hill \textsuperscript{1,4}\kern-1.5pt
\and Shuay-Pwu Patty Ho \textsuperscript{6}\kern-1.5pt
\and Johannes Hubmayr \textsuperscript{9}\kern-1.5pt
\and Brian Keating \textsuperscript{3}\kern-1.5pt
\and Kenji Kiuchi \textsuperscript{5,7}\kern-1.5pt
\and Akito Kusaka \textsuperscript{4,5,7}\kern-1.5pt
\and Adrian T Lee \textsuperscript{1,4}\kern-1.5pt
\and Michael Ludlam \textsuperscript{1}\kern-1.5pt
\and Aashrita Mangu \textsuperscript{1}\kern-1.5pt
\and Frederick Matsuda \textsuperscript{5}\kern-1.5pt
\and Heather McCarrick \textsuperscript{6}\kern-1.5pt
\and Federico Nati \textsuperscript{13}\kern-1.5pt
\and Michael D. Niemack \textsuperscript{14}\kern-1.5pt
\and Haruki Nishino \textsuperscript{7}\kern-1.5pt
\and John Orlowski-Scherer \textsuperscript{8}\kern-1.5pt
\and Mayuri Sathyanarayana Rao \textsuperscript{4}\kern-1.5pt
\and Christopher Raum \textsuperscript{1}\kern-1.5pt
\and Yuki Sakurai \textsuperscript{5}\kern-1.5pt
\and Maria Salatino \textsuperscript{10,11}\kern-1.5pt
\and Trevor Sasse \textsuperscript{1}\kern-1.5pt
\and Joseph Seibert \textsuperscript{3}\kern-1.5pt
\and Carlos Sierra  \textsuperscript{12}\kern-1.5pt
\and Maximiliano Silva-Feaver \textsuperscript{3}\kern-1.5pt
\and Jacob Spisak \textsuperscript{3}\kern-1.5pt
\and Sara M Simon \textsuperscript{12}\kern-1.5pt
\and Suzanne Staggs \textsuperscript{6}\kern-1.5pt
\and Osamu Tajima \textsuperscript{2}\kern-1.5pt
\and Grant Teply \textsuperscript{3}\kern-1.5pt
\and Tran Tsan \textsuperscript{3}\kern-1.5pt
\and Edward Wollack \textsuperscript{15}\kern-1.5pt
\and Bejamin Westbrook \textsuperscript{1}\kern-1.5pt
\and Zhilei Xu \textsuperscript{8}\kern-1.5pt
\and Mario Zannoni \textsuperscript{13}\kern-1.5pt
\and Ningfeng Zhu \textsuperscript{8}\kern-1.5pt}

\institute{\footnotesize
\noindent\textsuperscript{*} Corresponding author: \email{aamir.m.ali@berkeley.edu}\\
\noindent\textsuperscript{1}Department of Physics, University of California - Berkeley, Berkeley, CA\\ 
\noindent\textsuperscript{2}Kyoto University, Kyoto, Japan\\
\noindent\textsuperscript{3}University of California - San Diego, San Diego, CA\\
\noindent\textsuperscript{4}Lawrence Berkeley National Laboratory, Berkeley, CA\\
\noindent\textsuperscript{5}Kavli Institute for the Physics and Mathematics of the Universe, Tokyo, Japan\\
\noindent\textsuperscript{6}Princeton University, Princeton, NJ\\
\noindent\textsuperscript{7}University of Tokyo, Tokyo, Japan\\
\noindent\textsuperscript{8}University of Pennsylvania, Philadelphia, PA\\
\noindent\textsuperscript{9}National Institute of Standards and Technology, Boulder, Co\\
\noindent\textsuperscript{10}Stanford University, Stanford, CA\\
\noindent\textsuperscript{11}Kavli Institute for Particle Astrophysics and Cosmology, Stanford, CA\\
\noindent\textsuperscript{12}University of Michigan, Ann Arbor, MI\\
\noindent\textsuperscript{13}University of Milano - Bicocca, Italy\\
\noindent\textsuperscript{14}Cornell University, Ithaca, NY\\
\noindent\textsuperscript{15}NASA Goddard Spaceflight Center, Greenbelt, MD\\
}

\authorrunning{Aamir M. Ali et al.}
\titlerunning{Small Aperture Telescopes for the Simons Observatory}

\maketitle

\begin{abstract}
The Simons Observatory (SO) is an upcoming cosmic microwave background (CMB) experiment located on Cerro Toco, Chile, that will map the microwave sky in temperature and polarization in six frequency bands spanning 27 to 285 GHz. SO will consist of one 6-meter Large Aperture Telescope (LAT) fielding  $\sim$30,000 detectors and an array of three 0.42-meter Small Aperture Telescopes (SATs) fielding an additional 30,000 detectors. This synergy will allow for the extremely sensitive characterization of the CMB over angular scales ranging from an arcmin to tens of degrees, enabling a wide range of scientific output. Here we focus on the SATs targeting degree angular scales with successive dichroic instruments observing at Mid-Frequency (MF: 93/145 GHz), Ultra-High-Frequency (UHF: 225/285 GHz), and Low-Frequency (LF: 27/39 GHz). The three SATs will be able to map $\sim$10\% of the sky to a noise level of $\sim$\SI{2}{\micro\kelvin}-arcmin when combining 93 and 145 GHz. The multiple frequency bands will allow the CMB to be separated from galactic foregrounds (primarily synchrotron and dust), with the primary science goal of characterizing the primordial tensor-to-scalar ratio, $r$, at a target level of $\sigma \left(r\right) \approx 0.003$.


\end{abstract}

\section{Introduction}

Since its discovery in 1964~\cite{Penzias1965a} the CMB has been a key experimental window into the structure of the universe as a whole, and the early universe in particular. CMB measurements to date are consistent with an inflationary period in the very early universe~\cite{Baumann2009a}, but a predicted stochastic background of tensor perturbations (Inflationary Gravitational Waves) has yet to be detected. The key observable is a degree-scale signature in the divergence-free `B-mode' component of the CMB polarization~\cite{Seljak1996, Kamionkowski1997} parameterized by an overall amplitude given by the tensor-to-scalar ratio $r$, which probes the energy scale of inflation~\cite{Kamionkowski2016}. A number of experiments have attempted to detect the primordial B-mode signal to determine $r$.

This measurement is challenging. The B-mode signal is much fainter than the curl-free `E-mode' component, and considerably fainter than the polarized galaxy which is everywhere brighter than the primordial B-mode~\cite{Hinshaw2009}, with emission from synchrotron dominant at lower frequencies and from spinning dust dominant at higher frequencies~\cite{Krachmalnicoff2015}. The E-mode signal can also be gravitationally deflected by large-scale structure into a lensing B-mode signal which peaks at angular scales of $\ell \sim 1000$~\cite{Hanson2013, ThePOLARBEARCollaboration2014} ($\sim$ arcmins). Although scientifically interesting in their own right, the lensing B-modes act as a foreground to primordial B-modes. The primordial B-modes peak at much larger angular scales of $\ell\sim 90$ (degree scales, corresponding to the epoch of recombination) but the lensing B-modes dominate even at $\ell = 90$ if $r << 0.01$ \footnote{The primordial B-mode has a second peak at $\ell \lesssim 10$ corresponding to the epoch of reionization. These very large angular scales are not typically accessible from the ground, although this peak is a primary science target of possible upcoming CMB satellites, including the recently selected JAXA-led LiteBIRD satellite~\cite{Hazumi2019}, and the ground-based CLASS~\cite{Watts2018} and GroundBIRD~\cite{Oguri2016} experiments.}. In addition, recovering the larger angular scale modes where the primordial B-mode spectrum peaks requires excellent instrument stability over sufficiently long time scales (usually quantified in terms of $1/f$ noise). Lastly, spurious instrument effects or analysis inaccuracies can leak signal into the B-mode polarization~\cite{Lewis2001, Bunn2003}; a successful experiment thus requires combining excellent raw sensitivity with robust control of systematic errors.

\section{Experimental Strategy}
\begin{table}[]
\begin{tabular}{|lc|cccc|ll|}
\hline
\textbf{Focal Plane} & \textbf{\begin{tabular}[c]{@{}c@{}}Freq.\\ (GHz)\end{tabular}} & \textbf{$\boldsymbol{n_{\text{TES}}}$} & \textbf{\begin{tabular}[c]{@{}c@{}}$\boldsymbol{\Omega}_{Beam}$\\ (arcmin)\end{tabular}} & \textbf{\begin{tabular}[c]{@{}c@{}}Obs. Length\\ (Years)\end{tabular}} & \textbf{\begin{tabular}[c]{@{}c@{}}Map Depth\\ (\si{\micro\kelvin}$\cdot$arcmin)\end{tabular}} & \multicolumn{2}{c|}{\textbf{All SATs}}                 \\ \hline
\textbf{LF}          & \textbf{27}                                                        & 1050                                   & 91                                                                                       & 1                                                                      & 35                                                                                              & \textbf{FOV}                              & 35$^\circ$ \\
\textbf{LF}          & \textbf{39}                                                        & 1050                                   & 63                                                                                       & 1                                                                      & 21                                                                                              & \textbf{Aperture}                         & 42 cm      \\
\textbf{MF 1 \& 2}   & \textbf{93}                                                        & 23000                                  & 30                                                                                       & 5                                                                      & 2.6                                                                                             & \textbf{$\boldsymbol{f}$-\#}              & $\sim$1.6  \\
\textbf{MF 1 \& 2}   & \textbf{145}                                                       & 23000                                  & 17                                                                                       & 5                                                                      & 3.3                                                                                             & \textbf{$\boldsymbol{T}_{\text{CHWP}}$}   & 50 K       \\
\textbf{HF}          & \textbf{225}                                                       & 12000                                  & 11                                                                                       & 5                                                                      & 6.3                                                                                             & \textbf{$\boldsymbol{T}_{\text{Optics}}$} & 1 K        \\
\textbf{HF}          & \textbf{280}                                                       & 12000                                  & 11/9                                                                                     & 5                                                                      & 16                                                                                              & \textbf{$\boldsymbol{T}_{FP}$}            & 100 K      \\ \hline
\end{tabular}
\caption{Key SAT parameters. All numbers are approximate/projected. Frequencies given are nominal band centers. Number of TES detectors is total number of bolometers fielded, although the dichroic pixels share common antennae/optics. The 23000 MF detectors are shared evenly across the two MF instruments. The beam size given is at full-width half-maximum. The observing length is the expected nominal survey length. The final survey sensitivity is given as a map depth in \si{\micro\kelvin}$\cdot$arcmin, assuming a 20\% observing efficiency, and the SO `baseline noise model' assumed in the SO forecasting study~\cite{Ade2019}.}
\label{tablev0}
\end{table}

The SO SAT experimental strategy targets maximizing sensitivity to the faint primordial B-mode signal while minimizing and/or controlling systematic effects which would limit the ability to discriminate the primordial B-mode signal from others. To distinguish galactic foregrounds from the CMB, there will be three SAT telescopes deployed using differing frequencies of dichroic detectors. Two Mid-Frequency (MF: 93 and 145 GHz) instruments will be deployed first to maximize sensitivity near the foreground minimum, the third an Ultra-High-Frequency (UHF: 225 and 285 GHz) instrument to characterize dust. An additional Low-Frequency (LF: 27 and 39 GHz) optics tube will be deployed for a year to characterize synchrotron (table \ref{tablev0}). Fast optics (42 cm aperture, 35$^\circ$ Field of View, FoV) maximize throughput to the 7-detector tile focal plane. The optics and focal plane are cooled to 1 K and 100 mK, respectively, to maximize sensitivity. In total, we anticipate the instruments to have a sensitivity of 2 \si{\micro\kelvin}-arcmin in combined 93 and 145 GHz bands over $\approx 10\%$ of the sky~\cite{Ade2019}. 

Each SAT will operate with a Cryogenic Half--Wave Plate (CHWP) polarization modulator to reconstruct the CMB polarization. Half--Wave Plate polarization modulation has been employed on past CMB telescopes and has been shown to suppress long-timescale drifts ($1/f$ noise) from the atmosphere~\cite{Kusaka2018,Takakura2017}. The use of a CHWP (vs. ambient temperature HWP) stabilizes and dramatically reduces load from the HWP, improving sensitivity. Extensive warm and cryogenic baffling and shielding of various types have also been implemented to prevent systematic contamination from far-sidelobe pickup (in particular from ground and sun), ghosting inside the instrument, excess loading, magnetic fields, and Radio Frequency (RF) pickup. The detailed instrument design follows this section. 

The SAT program also benefits from synergy with the SO Large Aperture Telescope~\cite{Galitzki2018} which will generate high-sensitivity maps covering the entire SAT patch at smaller angular scales. In combination with data from other large-scale-structure surveys (e.g. from Cosmic Infrared Background~\cite{Sherwin2015}) these data allow a reconstruction of the lensing potential, which can then be used to clean lensing B-modes from the SAT patch (`delensing,' see e.g.~\cite{Larsen2016}), improving sensitivity to primordial B-modes. The forecasted $\sigma(r) \approx 0.003$ mentioned above assumes a relatively modest delensing efficiency of $A_{lens} = 0.71$ (see ref.~\cite{Ade2019} for more details) and conservatively assumes no CMB delensing from the LAT. In addition to delensing, the synergy between the SAT and LAT will allow SO to perform foreground characterization and separation over a wide range of angular scales, improving the robustness of the foreground cleaning strategy. 
\subsection{Scan Strategy}

The SAT scan will cover $\sim$ 10\% of the sky, split between a northern and southern field. Most of the integration time will be focused on the southern field; the northern field will be observed when the southern field is inaccessible. The SAT scan strategy minimizes the sky patch area to maximize map depth, and is refined via the use of an `opportunistic scheduler,' an algorithm which schedules observation according to a rank of targets among tiles in the observing patches based on visibility to the instrument, sun/moon avoidance, and integration time\footnote{A more detailed discussion of the opportunistic scheduler can be found in Stevens et al. 2018.~\cite{StevensJR2018}}. 

\begin{figure}[t]
\begin{center}
\includegraphics[width=0.95\linewidth, keepaspectratio]{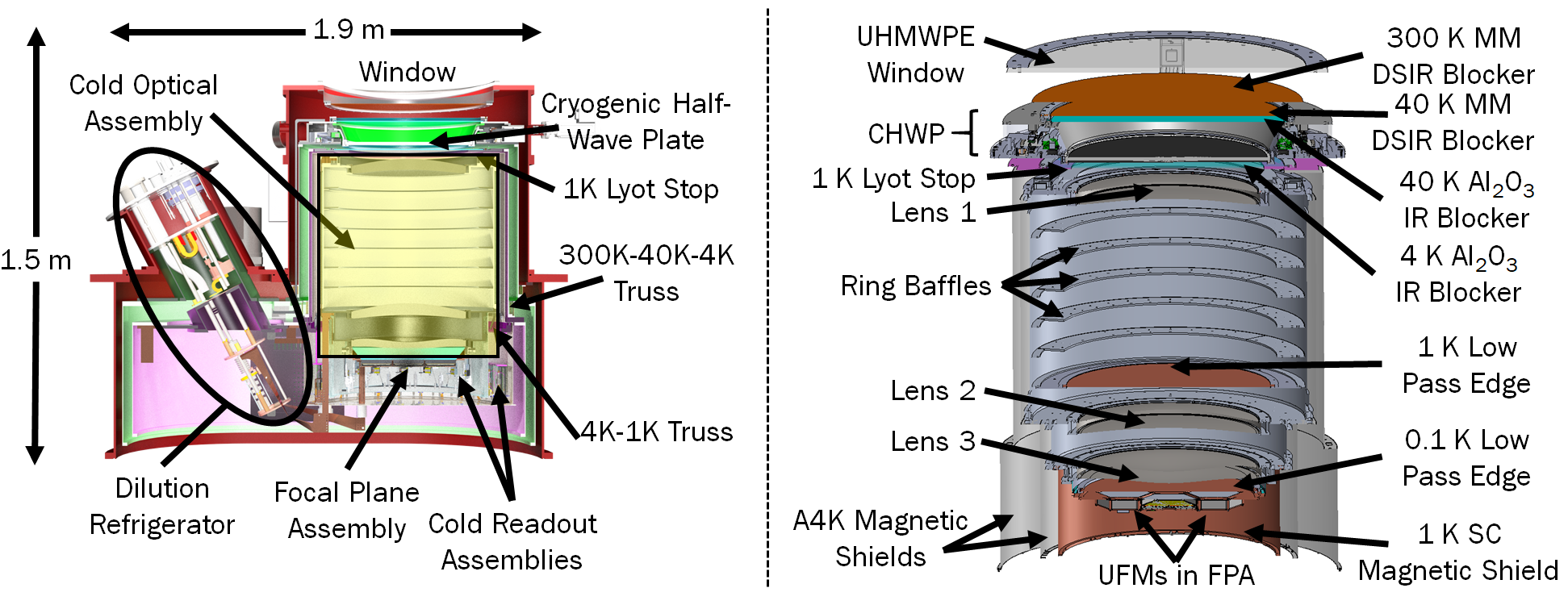}
\caption{{\it Left:} Simplified CAD render of SAT. The SAT receiver houses all major components, with an ambient temp. vacuum shell (red) enveloping PTC cooled stages at $\sim 40$ K (green, houses CHWP), $\sim 4$ K (purple, houses one A4K magnetic shield), and DR cooled stages at $\sim1$ K (houses COA, second A4K magnetic shield) and $\sim 100$ mK (houses focal plane/detectors). Readout wiring and components are distributed across all temperature stages. The 40 K and 4 K stages mount to the 300 K shell via concentric Al rings separated by G10 tabs (Figure \ref{sat_chwp} top right), while the 4 K, 1 K, and 100 mK links are made with pultruded carbon fiber trusses. {\it Right:} Closer view of COA and associated optical elements. The Ultra-High-Molecular-Weight Polyethylene (UHWMPE) window illuminates the CHWP, sky-side of the image forming 1 K Lyot Stop (first COA element). Three Si lenses refract light onto the detector UFMs held at 100 mK in the focal plane assembly, surrounded by a superconducting magnetic shield. Filters are implemented at every temperature stage to suppress out-of-band loading (Section \ref{sec:Cryogenic Receiver}).}
\label{fig:sat_coa}
\end{center}
\end{figure}

\section{Instrumental Design}

Each SAT is a compact 3-lens refractor with almost all major components housed in or attached to the cryogenic receiver. The SAT receiver relies on a pulse-tube cooler (PTC) and dilution refrigerator (DR) to cool the focal plane of transition--edge sensor (TES) detectors to 100 mK and the refracting optics to 1 K. The major subsystems are described below.

\subsection{Cryogenic Half Wave Plate}
\label{sec:chwp}
The SAT CHWP closely follows the mechanical design and optical properties of an analogous CHWP being deployed for the POLARBEAR 2b/c experiments~\cite{Hill2018}. 
The CHWP consists of a 3-layer birefringent sapphire crystal mounted in a rotor assembly which incorporates NdFeB magnets in a G10 ring to create a highly uniform permanent magnet ring. The rotor is suspended over a stator ring of superconducting YBCO (a type II superconductor with transition $\sim$ 90 K) via flux-pinning. A ring of synchronous solenoid motors on the stator creates a rotating magnetic field which couples to 80 alternating polarity NdFeB permanent magnet `sprockets' to drive the CHWP at a frequency of $\sim$2 Hz (8 Hz polarization modulation). The rotation of the CHWP is exceptionally smooth due to the combination of the magnetic levitation bearing and the contactless motor. An optical encoder provides feedback for the motor and encodes the angle of CHWP for incident polarization-angle reconstruction in the ensuing analysis. A full characterization of the stability of the CHWP rotation and of the resulting encoder signal will be presented in a forthcoming publication. We expect the CHWP temperature to be $\sim50$ K, and since the levitating rotor is only thermally coupled to its surroundings radiatively, significant temperature changes will only occur on timescales much longer than the duration of a single scan of the sky. Calibration of the polarization modulation will be achieved by periodically observing unpolarized sources through a sparse wire grid mounted immediately skyward of the window (Figure \ref{fig:det_mag_satp}).  

\subsection{Optics}

The SAT optics are designed to maximize throughput on the large focal plane  (seven 6-inch hexagonal detector array tiles) with the physical constraints set by mounting the 40 K CHWP sky-side of the refracting elements. The CHWP diameter is driven by the available size of sapphire (510 mm), and the SAT refractor consists of three metamaterial antireflection coated~\cite{Coughlin2018} Silicon lenses ($\sim3$-5 cm thickness) of 46 cm diameter, the largest Si lenses used for a CMB telescope to date. Respecting mounting considerations, the Si and Sapphire diameters set the Stop Aperture to 42 cm. The window to focal plane length is compact at 110 cm, and the optics are held in a stand-alone Cold Optical Assembly (COA, figure \ref{fig:sat_coa}) entirely cooled to $\sim1$~K by the DR, with the 1K Lyot Stop the first COA element. The optical design has a FoV of 35~$^\circ$, with a 93 GHz beam full-width at half-maximum (FWHM) of $\sim30$~arcmin, $f$-number of $\sim 1.6$ and is diffraction limited.

\begin{figure}[t]
\begin{center}
\includegraphics[width=0.95\linewidth, keepaspectratio]{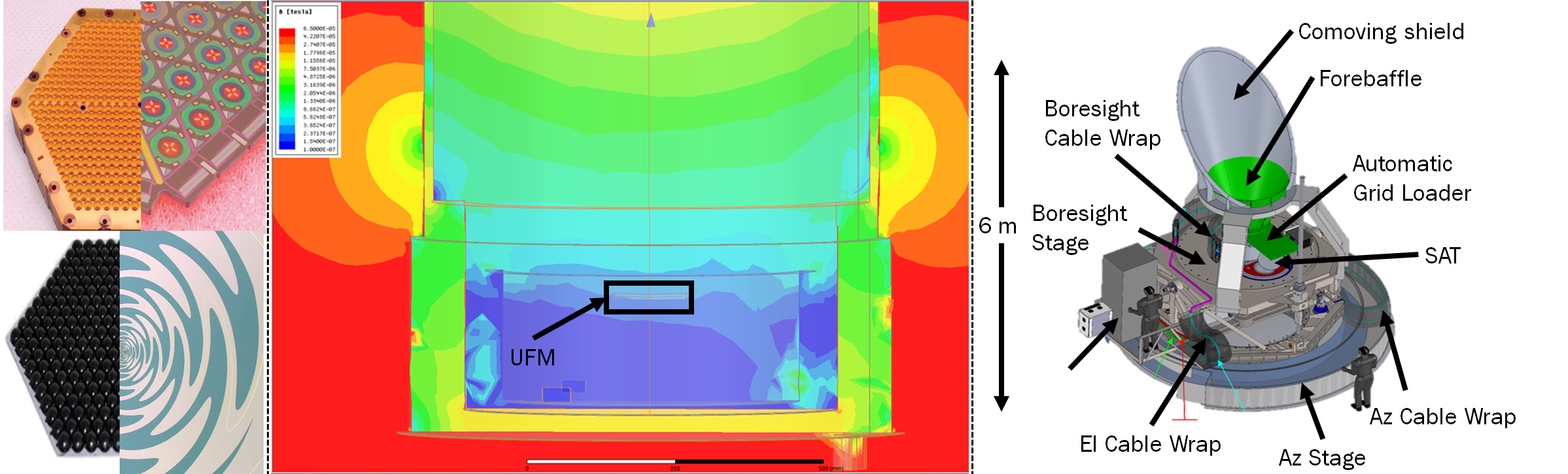}    
\caption{{\it Top Left:} View of Spline-profiled feedhorns next to OMT coupled NIST style detector array. {\it Bottom Left:} AR coated lenslet array next to zoom of half of one Berkeley style sinuous antenna. {\it Center:} Simulation of magnetic field in presence of three magnetic shield scheme described in section \ref{sec:RF and Magnetic Shielding}. The location of the UFM is shown, in a region with shielding $\sim$ 250x. {\it Right:} CAD render of SAT-platform (SATP). The SATP provides 3-axis actuation (Az 0-360$^\circ$, El 0-90$^\circ$, Deck $\pm 90^\circ$), houses utilities, cable wraps, and a co-moving shield attached to the elevation stage. A forebaffle (green) is attached to the front of the cryostat with an integrated automatic sparse-wire grid loader to be used intermittently for instrument calibration.}  
\label{fig:det_mag_satp}
\end{center}
\end{figure}

The SAT uses extensive baffling to control spurious loading. A free standing ground shield ($\sim$ 7 m radius, $\sim$ 5.5 m height), a comoving shield ($\sim$ 2.2 m, nominally reflective) mounted to the elevation stage, and a forebaffle ($\sim$1.7 m, nominally absorptive) attached to the front of the SAT  will all be implemented external to the cryostat to limit spillover systematic errors (figure \ref{fig:det_mag_satp}), in particular blocking light from the terrain and nearby mountains from entering into the SAT, and suppressing ground synchronous signal. Ring baffles $\sim$75 mm deep along the body of the COA between Lens 1 and 2 work in concert with a full COA interior coating of black tiles to absorb internal cold spillover.


\subsection{Cryostat}
\label{sec:Cryogenic Receiver}
The cryostat design was driven by the requirement to house/cool the optics/detectors and accommodate the insertion of the refrigerators at an angle. The 27.5$^\circ$ tilt allows the cryostat to rotate in boresight (BS) deck angle $\pm 75^\circ$ while keeping the PTCs within the required operating tilt of $<40^\circ$ at lower observing elevations. As such, the SAT focal plane should be able to image \textit{Taurus-A}, a supernova remnant in the Crab Nebula, which is an important on-sky polarization angle calibrator~\cite{Aumont2018}. 

The SAT cryostat\footnote{Manufactured by Criotec Impianti S.p.A., Chivasso, Italy} is cooled by a Bluefors\footnote{Bluefors Cryogenics, Helsinki, Finland} SD-400 DR providing continuous cooling at 1 K and 100 mK stages, and two PT-420\footnote{Cryomech Inc., Syracuse, NY} (one integrated with the DR assembly) PTCs providing 40 K and 4 K cooling stages. Each stage acts as a radiation shield for colder stages. The 40 K stage holds the CHWP assembly and cools the superconducting bearing (Section \ref{sec:chwp}) below its critical temperature ($T_C$) of $\sim$ 90 K. The 4 K stage houses one of two high-$\mu$ magnetic shields, while the entire COA and a second high-$\mu$ magnetic shield are cooled to 1 K. The detectors and the microwave SQUID multiplexing ($\mu$Mux) cold readout assembly are mounted in Universal Focal-plane Modules (UFM) held in a Focal Plane Assembly (FPA) at 100 mK. The RF signals for the $\mu$Mux readout  are brought to/from the UFMs with RF components and wiring distributed across the temperature stages (see Silva-Feaver et al. 2019~\cite{Silva-Feaver2019}). Experiments utilizing similar cryogenics~\cite{Iuliano2018} have achieved extremely high ($>$90\%) operating efficiency in terms of cryogenic up-time (i.e. excluding weather, data cuts, etc.).

An extensive filtering scheme along the optical path drastically cuts the out-of-band loading on the focal plane. Double-sided IR Metal Mesh (MM DSIR) filters reflect out as much incident radiation as possible. Alumina IR blocking filters absorb what makes it through and conduct the absorbed heat away, reradiating at a much lower temperature. Low-Density Polyethylene Low Pass Edge (LPE)\footnote{We note here that the LPE's are not band defining filters; band definition takes place via on-chip filtering integrated into the detector circuit.} filters prevent high frequency blue-leak from coupling into the detectors. All filters, as well as the 680 mm Ultra-High-Molecular-Weight Polyethylene (UHMWPE) window, are AR coated.

\subsection{Detectors and Readout}
The Simons Observatory employs superconducting aluminum-manganese TES bolometers tuned to a critical temperature of $T_C \sim 160$ mK, comfortably above the expected operating temperature of the DR cold stage. The detector noise is below the fundamental photon noise lower limit (i.e. they are background limited). In order to improve sensitivity, SO uses arrays of TESs to significantly increase detector count, deploying two flavors of arrays of antenna-coupled TESs, sinuous antennas coupled via lenslet arrays~\cite{Beckman} for the LF detectors, and orthomode-transducers (OMTs) coupled via feedhorn arrays~\cite{Simon2018} for the MF and HF detectors (Figure \ref{fig:det_mag_satp}). The use of two technologies enables efficient parallel fabrication paths, allows for independent checks of systematics peculiar to each, and allows for the advancement of both detector technologies for future experiments.

The TES bolometers are read out with $\mu$Mux SQUID multiplexing architecture~\cite{Irwin2004}, in which each DC biased TES is coupled to an rf-SQUID that inductively loads a high-Q microwave resonator. As the TES current fluctuates in response to the photon signal, the resonant frequency of the resonator is modulated. Resonators of varying resonant frequencies are coupled to a microwave transmission line, and the resonance fluctuation is interrogated by probe tones. Flux-ramp modulation along a second DC bias line linearizes the SQUID response and modulates the science signal. The two bias lines and one microwave transmission line are all the wiring required for one multiplexing chain. The goal is for a single readout line to multiplex the 1764 bolometers found in one MF or UHF detector array.

The feedhorn/lenslet array, the detector array, and an assembly containing the 100 mK $\mu$Mux readout components called the Universal $\mu$Mux Module constitute a modular UFM. Seven UFMs are mounted in a focal plane assembly (FPA) which mounts the detectors a precise distance from the third lens to optimize throughput and Strehl ratio. The FPA is surrounded by a series of concentric rings at different temperatures that mount readout components, wiring, and accomplish the transition between temperature stages: one assembly bridging 100~mK to 1~K, and another  1~K to 4~K~\cite{Silva-Feaver2019}. Isothermal cables from this ring run to a readout harness which spans the 4~K to 40~K to 300~K stages and houses the remaining internal readout components and wiring. The warm readout is handled by the SLAC microresonator radio frequency electronics system (SMuRF~\cite{Henderson2018}). The SAT FPA is designed for modularity, allowing individual UFMs to swap easily without disassembling the readout, and allowing for easy swapping of readout components along an individual readout chain, as necessary.

\subsection{RF and Magnetic Shielding}
\label{sec:RF and Magnetic Shielding}
To prevent spurious RF and magnetic pickup, a comprehensive shielding strategy has been developed for the SAT. Telescope slew through the geomagnetic field can introduce a ground synchronous signal, primarily by coupling to the SQUID readout. To mitigate this the SAT has 3 magnetic shields: two high-permeability\footnote{material A4K, shields manufactured by Amuneal Corp., Philadelphia, PA} shields along the length of the cryostat, and an additional superconducting shield surrounding the focal plane. In combination, the shielding scheme is simulated to provide better than a factor of 200 suppression of the DC magnetic field at the SQUIDs (figure \ref{fig:det_mag_satp}), which we calculate to be below their sensitivity.\footnote{The magnetic shielding strategy will be discussed at greater length in an upcoming publication.}

Although the environment is relatively RF quiet, any spurious RF signal can couple to the detectors/readout. To mitigate this, the lower portion of the vacuum shell of the cryostat serves as a Faraday cage. Breaks in the electrical continuity are bridged as follows: all connector boxes mounting to the cryostat are RF tight, all wiring in and out of the cryostat is shielded, all vacuum flanges are fitted with conductive gasket seals, and the conductive breaks across thermal stages are bridged with a thin layer of aluminized mylar (Figure \ref{sat_chwp} Top Right). Electrical continuity from the FPA to the detector ground planes completes the cage. During vacuum evacuation and venting, the front and back volumes of the cryostat (either side of the aluminized mylar) are externally linked in order to prevent a pressure gradient across the aluminized mylar.

\subsection{Platform}
The SAT platform\footnote{Designed and manufactured by Vertex Antennentechnik GmbH, Duisburg, Germany} (Figure \ref{fig:det_mag_satp})  provides 3-axis actuation in Azimuth (0-360$^\circ$), Elevation (0-90$^\circ$), and along the Boresight axis ($\pm 90 ^\circ$, sometimes referred to as `deck angle') The additional third axis changes the angle of the SAT with respect to the sky, a powerful insight into systematic effects and a crucial data split for null tests of biases in the analysis pipeline~\cite{BICEP2Collaboration2015}. This will also be the first demonstration of using both a CHWP and boresight rotation to constrain systematic errors. 
  
\section {Current Status and Conclusions}
\begin{figure}[t]
\begin{center}
\includegraphics[width=0.95\linewidth, keepaspectratio]{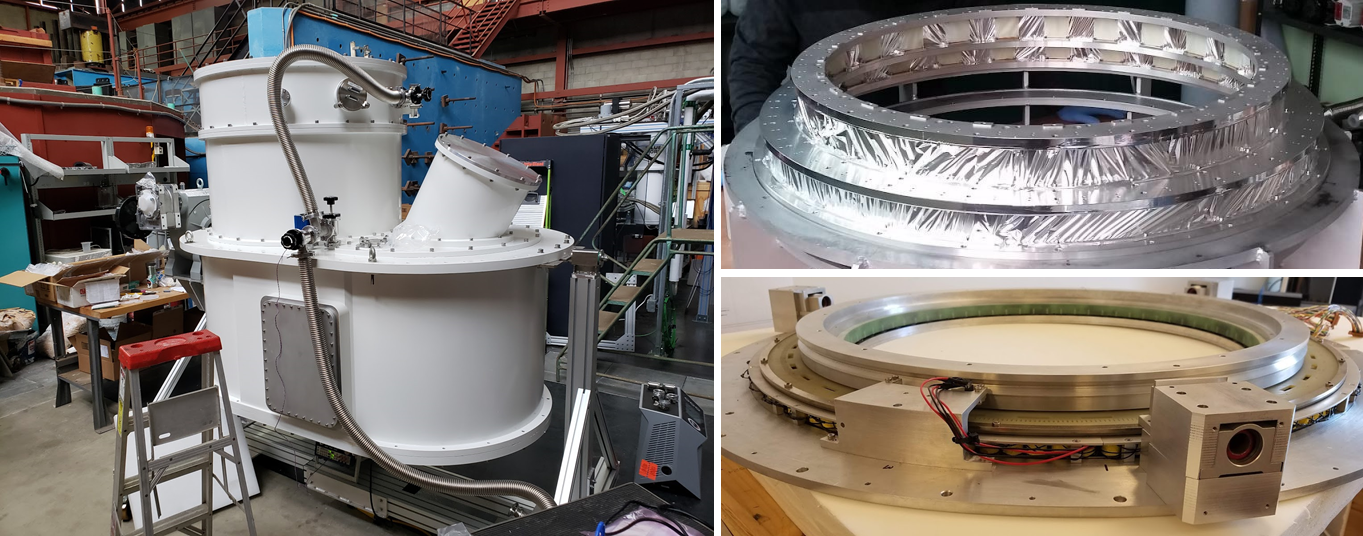}
\caption{{\it Left:} SAT-1 in high-bay at the University of California-San Diego, where it is presently undergoing cryogenic acceptance tests. {\it Top Right:} Picture of the assembly which mechanically mounts and thermally isolates the 300 K, 40 K, and 4 K stages of the SAT. G10 tabs link the Al-6061 rings. A layer of aluminized mylar (visible from the outside) shorts the adjacent stages electrically and completes a Faraday cage around the detectors for RF shielding. {\it Bottom Right:} Largely completed CHWP at Lawrence Berkeley National Laboratory, where it is presently undergoing validation.}
\label{sat_chwp}
\end{center}
\end{figure} 

The first SAT cryostat has already been received, and many of the major subsystems (COA, CHWP, FPA, readout assemblies, etc.) are in advanced stages of completion (figure \ref{sat_chwp}). Integration and testing of the first MF SAT will continue through mid 2020, with an expected SAT 1 first light in Chile in late 2020/early 2021. Since the SAT program is highly parallelized, the second and third SAT receiver and most associated subsystems will arrive by mid 2020. After SAT-1 is deployed, we expect a rapidly phased deployment of the remaining SATs and full observatory operation by 2022. 

A detailed forecasting campaign has been conducted (and described in detail in the recent SO forecasting paper~\cite{Ade2019}) for both the LAT and SATs under two sets of assumptions regarding the noise model of the observatory: a nominal `baseline' level assuming only modest technical advancement over existing experiments, and a more aggressive `goal' level. Under the baseline model, presuming a nominal 5 year survey (1 year for LF) with an assumed 20\% total observing efficiency\footnote{consistent with the realized performance of the POLARBEAR and ACT experiments}, the SATs are forecasted to map $\sim$10\% of the sky to a noise level of $\sim$2 \si{\micro\kelvin}-arcmin when combining 93 and 145 GHz (table 1); at these map-depths with foreground cleaning from the multiple frequency bands, the SATs are projected to characterize $r$ to a target level of $\sigma \left(r\right) \approx 0.003$.

\begin{acknowledgements}
This work was supported in part by a grant from the Simons Foundation (Award \# 457687.)
\end{acknowledgements}

\pagebreak

\bibliography{report.bib} 

\end{document}